\documentclass[a4paper,12pt]{article}

\usepackage{amsmath,amssymb,booktabs}
\usepackage{epsf,epsfig}
\usepackage{psfrag}
\usepackage{cite}
\usepackage{mathrsfs}
\usepackage{bbm}
\usepackage[caption=false]{subfig}
\usepackage{dcolumn}
\usepackage[export]{adjustbox}
\allowdisplaybreaks

\def\be{\begin{equation}}
\def\ee{\end{equation}}
\def\beq{\begin{eqnarray}}
\def\eeq{\end{eqnarray}}

\def\DB0{\partial B_0}

\def\Cl2{\mbox{Cl}_2}

\def\Eqref#1{(\ref{#1})}

\def\s{\hat s}

\usepackage{color}

\definecolor{Brown}{rgb}{0.5,0.25,0}

\begin{document}


\begin{titlepage}

\begin{flushright}
SI-HEP-2018-22\\
QFET-2018-13\\[0.1cm]
\today
\end{flushright}
\vskip 1.0cm

\begin{center}
\Large{\bf\boldmath
Five-particle contributions to the \\ inclusive rare $\bar B \to X_{s(d)} \, \ell^+\ell^-$ decays
\unboldmath}

\normalsize
\vskip 1.5cm

{\sc Tobias~Huber},
{\sc Qin~Qin}
and
{\sc K.~Keri~Vos}\\

\vskip 1.5cm

{\it Theoretische Physik 1, Naturwissenschaftlich-Technische Fakult\"at,
\\ Universit\"at Siegen, Walter-Flex-Strasse 3, D-57068 Siegen, Germany}

\vskip 2.1cm

\end{center}

\begin{abstract}
\noindent
We calculate tree-level contributions to the inclusive rare $\bar B \to X_{s(d)} \, \ell^+\ell^-$ decays.
At the partonic level they stem from the five-particle process $b \to s(d) \, q \bar q \, \ell^+\ell^-$, with $q \in \{u,d,s\}$.
While for $b \to d$ transitions such five-body final states contribute at the same order in the Wolfenstein expansion compared to the three-body partonic decay, they are CKM suppressed in $b \to s$ decays. In the perturbative expansion, we include all leading-order contributions, as well as partial next-to-leading order QCD and QED effects.
In the case of the differential branching ratio, we present all results
completely analytically in terms of polylogarithmic functions of at most weight three. We also consider the
differential forward-backward asymmetry, where all except one interference could be obtained analytically.
From a phenomenological point of view the newly calculated contributions are at the percent level or below.
\end{abstract}

\vfill

\end{titlepage}


\section{Introduction}
\label{sec:intro}

Despite tremendous effort, the experiments at the LHC have to date not found any evidence for a direct signal of physics beyond the Standard Model (SM). In the absence of direct new-physics signals, indirect searches via low-energy observables become ever more important. In case the latter are suppressed or forbidden at tree level in the SM, they are capable of probing high scales via virtual effects of new degrees of freedom. Depending on the process, the scales probed in low-energy observables might even outrange those accessible via on-shell production.

As a matter of fact, all currently observed tensions between experimental data and SM predictions --~the so-called anomalies~-- are in low-energy observables located in the (quark or lepton) flavour sector of the SM. Among the most prominent ones, there are $P^\prime_5$ \cite{DescotesGenon:2012zf} in $\bar B \to K^\ast \ell^+\ell^-$, $R_{D^{(\ast)}}$, $R_{K^{(\ast)}}$, and $(g-2)_{\mu}$~\cite{Bennett:2006fi}. All except the last one stem from {\emph{exclusive}} decays of heavy mesons, and experimental measurements rely mostly on data from the $B$-factories \cite{Lees:2013uzd,  Lees:2012xj, Huschle:2015rga,Abdesselam:2016llu,Wehle:2016yoi}, LHCb \cite{Aaij:2014ora, Aaij:2017vbb, Aaij:2015oid, Aaij:2015yra}
, and partially ATLAS \cite{Aaboud:2018krd} and CMS \cite{Sirunyan:2017dhj}.

One useful way to shed light on the nature of the anomalies is a cross-check via the corresponding observables in the {\emph{inclusive}} modes. With the first data-taking at Belle~II in sight, there is a unique opportunity to get access to these modes with sufficient precision both on the theoretical and experimental side. On the theory side, obtaining precise predictions amounts to including higher-order perturbative corrections on the one hand, but also multi-particle contributions on the other. In the case of rare and radiative flavour-changing neutral current transitions the focus in the past was largely on $b \to s$ decays such as $\bar B \to X_s \gamma$ and $\bar B \to X_s \ell^+ \ell^-$, which have both reached a highly sophisticated level in terms of inclusion of perturbative corrections (for the latest comprehensive analyses, see~\cite{Czakon:2015exa,Misiak:2015xwa} and~\cite{Huber:2015sra}, respectively). While in $\bar B \to X_s \gamma$ four-body contributions which at the partonic level amount to $b \to s \, q \bar q \, \gamma$ (with a light quark $q \in \{u,d,s\}$) have been calculated to leading~\cite{Kaminski:2012eb} and next-to-leading order~\cite{Misiak:2010tk,Huber:2014nna}, the corresponding five-particle $b \to s \, q \bar q \, \ell^+\ell^-$ modes to $\bar B \to X_s \ell^+ \ell^-$ are yet unknown.

Besides $b \to s$ decays, also $b \to d$ transitions will become relevant in the Belle~II era. They are interesting on their own grounds,
because contrary to $b \to s$, all three sides of the unitarity triangle for $b \to d$ transitions are of the same order ${\cal O}(\lambda^3)$ in the Wolfenstein expansion parameter $\lambda \approx \left| V_{us} \right|$. This means in particular that matrix elements of the effective operators $P_{1,2}^u$ are not CKM suppressed compared to those of their $P_{1,2}^c$ siblings. On general grounds, one can therefore expect quite sizeable CP-violating effects in $b \to d$ modes.

While there are plenty of papers dealing with the inclusive $\bar B \to X_s \ell^+ \ell^-$ decay, there is only little dedicated literature on the $\bar B \to X_d \ell^+ \ell^-$ counterpart. The next-to-next-to-leading order (NNLO) corrections to the matrix elements have been computed in~\cite{Asatrian:2003vq,Seidel:2004jh}, and also the latest phenomenological study dates back fifteen years~\cite{Asatrian:2003vq}. In view of prospects on the experimental side, also a theory update is timely. We start this endeavour in the present article by computing the five-particle $b \to d \, q \bar q \, \ell^+\ell^-$ process with $q \in \{u,d,s\}$ at tree level. We compute these corrections for two distributions that are differential in the dilepton invariant-mass squared, namely the branching ratio and the forward-backward asymmetry. It turns out that the integrated matrix elements of the dimension-six operators are real-valued due to the absence of strong phases, and hence do not affect CP-violating observables. The corresponding contributions to $\bar B \to X_s \ell^+ \ell^-$ are obtained by a simple replacement of the CKM factors.

On the technical side we have to deal with the integration of the squared matrix elements over the five-particle massless phase space. While several parametrisations exist~\cite{Byckling:1971vca,Kumar:1970cr,Heinrich:2006sw} which all turn out to be useful for particular types of interferences (which we specify in subsequent sections) the analytic integration over the kinematic invariants turns out to be the main challenge since we aim at performing all integrations in an analytical manner. While many of the previous studies of processes with five final-state particles~\cite{Achasov:2003re, Pruna:2016spf,Fael:2016yle,Cata:2016epa} use numerical methods, our paper serves as a proof-of-principle that five-particle massless phase-space integrations can be carried out (almost) completely analytically in terms of polylogarithmic functions of at most weight three. Another recent work which evaluates the five-particle master integrals with massless particles analytically in dimensional regularisation was presented in~\cite{Gituliar:2018bcr}. That paper integrates the integral kernels over the entire phase space, whereas in the present paper we stay differential in the dilepton invariant-mass squared. This entails on the one hand that we do not encounter any infrared divergences in the present calculation. On the other hand, we do not have the freedom to exploit the full symmetry of the phase-space measure.

This article is organised as follows: In section two we specify the operator basis and explain the power-counting by means of which we decide which terms to include in the present calculation. Section three contains details on the phase-space integration, and in section four we present our results for the differential branching ratio and forward-backward asymmetry. We finally conclude in section five, including a numerical estimate of the impact of the five-particle contributions. In two dedicated appendices, we illustrate how to analytically integrate sample kernels over the five-particle phase space, and collect the functions that appear in our analytical results.

\section{Effective theory and power counting}
\label{sec:me}


The calculation of transition amplitudes in heavy-quark decays is most conveniently performed in an effective field theory where the top quark, the heavy gauge bosons and the Higgs field have been integrated out~\cite{Buchalla:1995vs}. In what follows we present the formulas relevant for $b \to d$ transitions. The corresponding formulas for $b \to s$ are obtained by obvious adjustments of quark fields and CKM factors.

\subsection{The effective theory}
\begin{figure}[t]
	\centering
	\subfloat[]{\label{fig:P16I} \includegraphics[height=0.2\textwidth, valign=t]{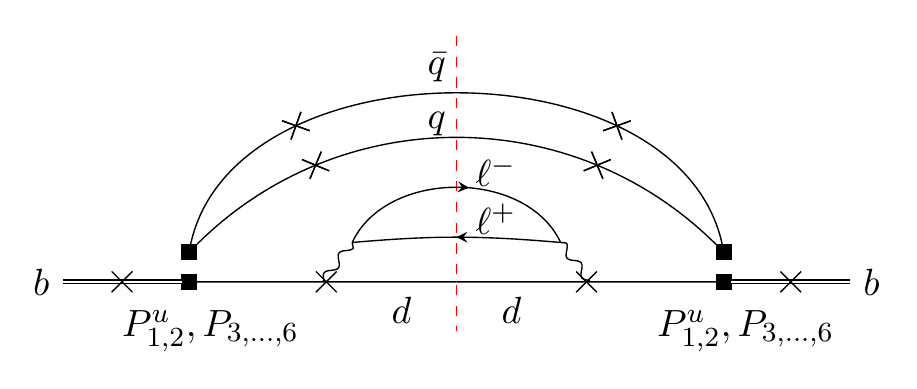}}
\subfloat[]{\label{fig:P16II} \includegraphics[height=0.2\textwidth, valign=t]{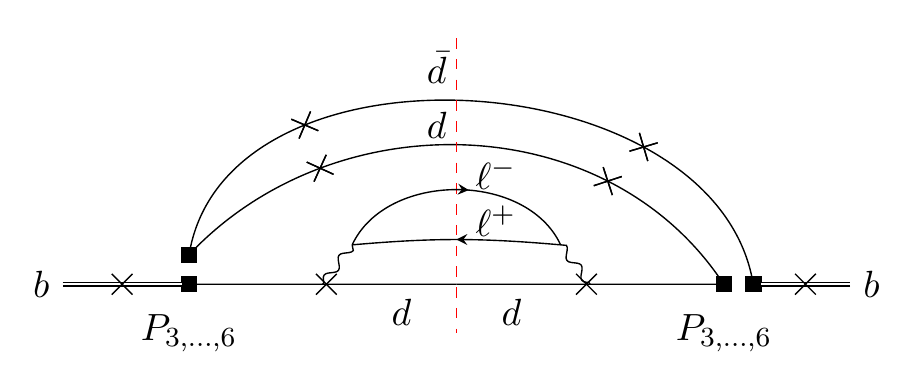}}
	\caption{Contributions of the operators $P_{1,2}^u,P_{3,\ldots,6}$ to the five-body decay, where the lepton pair can be emitted and absorbed from any of the crosses. Insertion (b) is only present for penguin operators $P_{3,\ldots,6}$ with $q=d$. }
	\label{fig:P16}
\end{figure}
The Lagrangian of the effective theory reads 
\newpage
\begin{align}\label{Leff}
\allowdisplaybreaks
\mathcal{L}_\text{eff} = & \mathcal{L}_{\text{QCD} \times \text{QED}}(u,d,s,c,b,e,\mu,\tau) \nonumber\\
&-\frac{4 G_F}{\sqrt{2}} \sum\limits_{p=u,c}V^*_{pd}V_{pb}(C_1 P_1^p + C_2 P_2^p)
+\frac{4 G_F}{\sqrt{2}} V^*_{td} V_{tb} 
\sum_{i=3}^{10} C_i(\mu) P_i,
\end{align}
where
\begin{align} \label{ope}
\begin{array}{rl}
P_1^p   = & (\bar{d}_L \gamma_{\mu} T^a p_L) (\bar{p}_L \gamma^{\mu} T^a b_L),
\vspace{0.2cm} \\
P_2^p   = & (\bar{d}_L \gamma_{\mu}  p_L) (\bar{p}_L \gamma^{\mu}     b_L),
\vspace{0.2cm} \\
P_3   = & (\bar{d}_L \gamma_{\mu}     b_L) \sum_q (\bar{q}\gamma^{\mu}     q),
\vspace{0.2cm} \\
P_4   = & (\bar{d}_L \gamma_{\mu} T^a b_L) \sum_q (\bar{q}\gamma^{\mu} T^a q),    
\vspace{0.2cm} \\
P_5   = & (\bar{d}_L \gamma_{\mu_1}
                     \gamma_{\mu_2}
                     \gamma_{\mu_3}    b_L)\sum_q (\bar{q} \gamma^{\mu_1} 
                                                         \gamma^{\mu_2}
                                                         \gamma^{\mu_3}     q),     
\vspace{0.2cm} \\
P_6   = & (\bar{d}_L \gamma_{\mu_1}
                     \gamma_{\mu_2}
                     \gamma_{\mu_3} T^a b_L)\sum_q (\bar{q} \gamma^{\mu_1} 
                                                            \gamma^{\mu_2}
                                                            \gamma^{\mu_3} T^a q),
\vspace{0.2cm} \\
P_7   = &  \frac{e}{16 \pi^2} m_b (\bar{d}_L \sigma^{\mu \nu}     b_R) F_{\mu \nu},
\vspace{0.2cm} \\
P_8   = &  \frac{g}{16 \pi^2} m_b (\bar{d}_L \sigma^{\mu \nu} T^a b_R) G_{\mu \nu}^a, 
\vspace{0.2cm} \\
P_9      = & (\bar{d}_L \gamma_{\mu} b_L) \sum_l (\bar{l}\gamma^{\mu} l),
\vspace{0.2cm} \\
P_{10}   = & (\bar{d}_L \gamma_{\mu}     b_L) \sum_l (\bar{l}\gamma^{\mu} \gamma_5 l),
\end{array} 
\end{align}
and $q=u, d, s, c, b$ and $l$ runs over the three charged lepton flavors. The five-particle contributions to $\bar B \to X_d \ell^+\ell^-$ amount at the partonic level to $b \to d \,  \, \ell^+ \ell^-  q \, \bar q$, where {\emph{by definition}} the final state is free of charm quarks, i.e.\ $q\in\{u,d,s\}$. At leading order in the perturbative expansion, one has to calculate interferences of the operators $P_{1,2}^u, \, P_{3, \ldots, 6}$ at tree level, which contribute at ${\cal O}(\alpha_e^2)$ as illustrated in Fig.~\ref{fig:P16}.  Here the lepton pair can be emitted and absorbed in the 16 different ways indicated by the crosses. For the penguin operators $P_{3, \ldots, 6}$ for $q=d$ in addition also the insertion in Fig.~\ref{fig:P16II} contributes.

From a technical point of view, one has to evaluate the squared matrix elements according to the interferences shown in Fig.~\ref{fig:P16}, and subsequently integrate them over the five-particle massless phase space. 
Since we stay differential in the dilepton invariant-mass squared, we do not encounter any infrared divergences and hence can perform the entire calculation in four-dimensional space-time. As a result, the interferences between the operators $P_{1,2}^u, \, P_{3, \ldots, 6}$ can be computed completely analytically.
When calculating the matrix elements, the following relations turn out to be useful,
\begin{align}
P_5 {}&= 10P_3 + 6(\bar{d}_L \gamma_{\mu}\gamma_5    b_L) \sum_q (\bar{q}\gamma^{\mu} \gamma_5    q), \ \\
P_6 {}&= 10P_4 + 6(\bar{d}_L \gamma_{\mu}\gamma_5   T^a b_L) \sum_q (\bar{q}\gamma^{\mu} \gamma_5  T^a  q) \, .
\end{align}
They hold in four dimensions and shorten the occurring traces over Dirac structures considerably.

\subsection{Power counting and higher orders}

\begin{figure}[t]
	\centering
	\subfloat[]{\label{fig:P27I} \includegraphics[height=0.2\textwidth, valign=t]{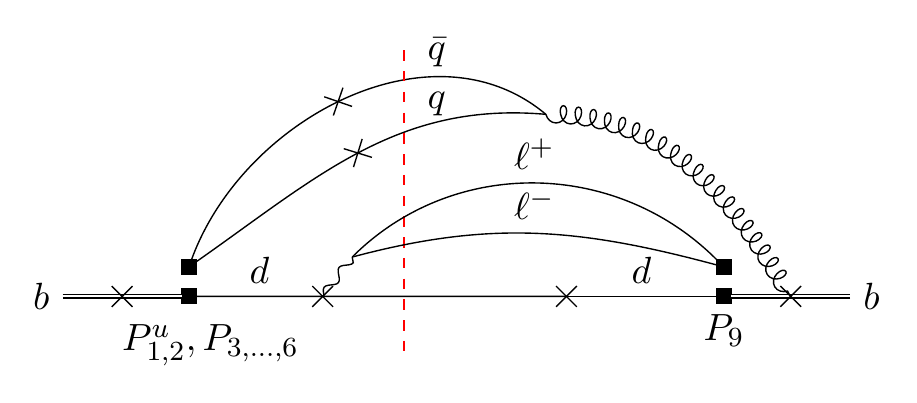}}
\subfloat[]{\label{fig:P27II} \includegraphics[height=0.2\textwidth, valign=t]{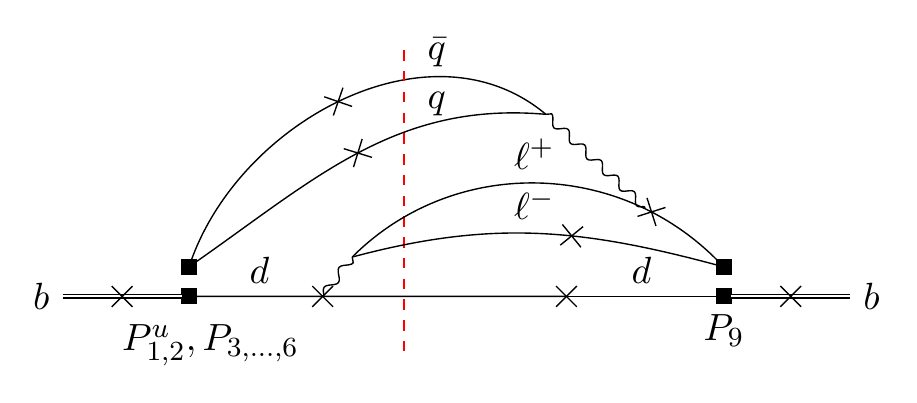}}
	\caption{Contributions of the $P_{1,2}^u, P_{3, \ldots,6}$ interference with $P_{9}$ via (a) emission of the quark pair via a gluon (QCD) (b) emission of the quark pair via a photon (QED). Again the photon or gluon can be emitted and absorbed from the different fermion lines as indicated by the crosses.}
	\label{fig:P27}
\end{figure}

In the presence of electroweak corrections, the perturbative expansion in inclusive $\bar B \to X_{s(d)}\ell^+\ell^-$ is consistently done in terms of $\widetilde\alpha_s = \alpha_s/(4\pi)$ and $\kappa = \alpha_e/\alpha_s$, as was pointed out in~\cite{Huber:2005ig}.
Following~\cite{Huber:2005ig,Huber:2007vv}, we only keep contributions up to $\mathcal{O}(\widetilde\alpha_s^3 \kappa^3)$ to the differential decay rate. As discussed above, the leading five-body contributions to $\bar B \to X_d \ell^+ \ell^-$ arise from the $P_{1,2}^u,P_{3,\ldots, 6}$ operators and are $\mathcal{O}(\widetilde\alpha_s^2\kappa^2)$.

In order to specify the set of interferences that we take into account beyond $\mathcal{O}(\widetilde\alpha_s^2 \kappa^2)$, we also consider possible suppressions by the Wilson coefficients. The numerical values for the Wilson coefficients can be found in Ref.~\cite{Huber:2005ig}. Through renormalisation group running, the Wilson coefficients $C_{1, \ldots, 8}(\mu_b)$ at a scale $\mu_b\sim m_b$ start to receive contributions at ${\cal O}(\widetilde\alpha_s^0\kappa^0)$, while $C_9$ and $C_{10}$ are given by\cite{Huber:2005ig} 
\begin{align}
C_9(\mu_b) & =3.7 \cdot 10^{-2} \kappa(\mu_b) + 1.9\widetilde\alpha_s(\mu_b)\kappa(\mu_b) + {\cal O}(\widetilde\alpha_s^2,\kappa^2) \, ,\\
C_{10}(\mu_b) & = 0.5 \cdot 10^{-2}\kappa^2(\mu_b)  -4.2\widetilde\alpha_s(\mu_b)\kappa(\mu_b) -3.8 \widetilde\alpha_s(\mu_b)\kappa^2(\mu_b) + {\cal O}(\widetilde\alpha_s^2,\kappa^3)
\end{align}
at $\mu_b = 5$ GeV. The Wilson coefficients $C_{9,10}$ will be counted as starting from $\mathcal{O}(\widetilde\alpha_s\kappa)$ due to the numerical smallness of their respective first coefficient.

Keeping the aforementioned counting of the Wilson coefficients in mind, we now turn to the operators $P_{7, \ldots, 10}$. At tree level they contribute by emitting a $q\bar{q}$-pair via a gluon or photon as illustrated in Fig.~\ref{fig:P27}. We label these emissions ``QCD'' and ``QED'', respectively. The interferences of $P_{7, \ldots, 10}$ with $P_{1,2}^u, \, P_{3, \ldots, 6}$ are then suppressed by least an additional coupling factor $\widetilde\alpha_s$ (QCD emission) or $\widetilde\alpha_s\kappa$ (QED emission) with respect to the pure $P_{1,2}^u, \, P_{3, \ldots, 6}$ interferences, while interferences between $P_{7, \ldots, 10}$ among themselves are all beyond ${\cal O}(\widetilde\alpha_s^3\kappa^3)$. In our analysis, we therefore only consider the interference of $P_{1,2}^u, P_{3,\ldots,6}$ with $P_{7, \ldots, 10}$.

We are aware of the fact that there are one-loop insertions from $P_{1, \ldots, 6}$ that start contributing at ${\cal O}(\widetilde\alpha_s^3\kappa^2)$, i.e.\ at the same order as the tree-level interferences involving the $P_{7, \ldots, 10}$ operators.
Including these loops would require performing our calculation in $D$ dimensions.
Since the aim of the present paper is to perform a first exploration of five-body contributions, inclusion of these loops is beyond the scope of the current analysis. To set the stage for a future calculation of five-body loop corrections, we find it nevertheless beneficial to take into account all tree-level contributions up to $\mathcal{O}(\widetilde\alpha_s^3 \kappa^3)$ and to give a numerical estimate of the impact of the higher-order terms. 

\begin{figure}[t]
	\centering
	\subfloat[]{\label{fig:P210I} \includegraphics[height=0.2\textwidth, valign=t]{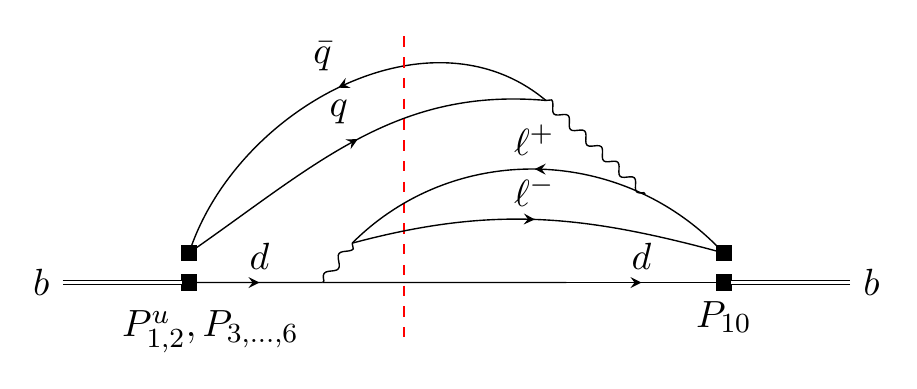}}
\subfloat[]{\label{fig:P210II} \includegraphics[height=0.2\textwidth, valign=t]{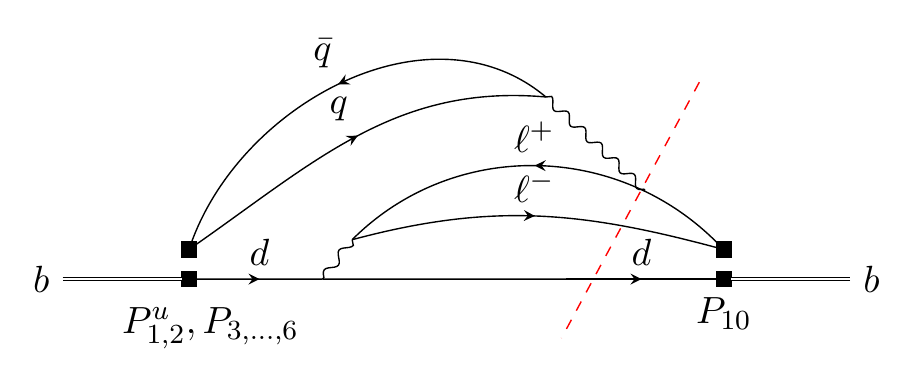}}
	\caption{Illustration of the $P_{1,2}^u, P_{3, \ldots,6}$ interference with $P_{10}$ where the $q\bar q$-pair is emitted from the charged leptons. The cancellation of infrared divergences in the five-particle cuts (a) requires fewer-particle cuts involving loops as exemplified in (b).}
	\label{fig:P210}
\end{figure}

As discussed, the $P_{7, \ldots, 10}$ operators contribute via a QCD or QED emission of the $q\bar{q}$-pair and interference with $P_{1,2}^u, P_{3, \ldots, 6}$ (see Figs.~\ref{fig:P27I} and~\ref{fig:P27II}).
While we take all QCD emissions into account, we treat the QED emissions as follows.
In case of the branching ratio, we include the full QED emission of the operators $P_{7,8,9}$. Here it turns out that the interferences where the $q\bar{q}$ pair is emitted from the lepton pair is zero for the $P_{7}$ and $P_9$ operators by symmetry of the phase-space measure. However, for the operator $P_{10}$, which involves the
axialvector current, these particular QED emissions do contribute and turn out to be infrared divergent. We illustrate them in Fig.~\ref{fig:P210I}. To cancel the divergent part of this contribution would require the inclusion of fewer-particle cuts involving loops such as the two-loop contribution of the three-body decay depicted in Fig.~\ref{fig:P210II}. As mentioned above, we presently only work at tree level and therefore omit the full QED emission of $P_{10}$ and leave it for future work. For the forward-backward asymmetry, the situation is reversed and actually the $P_{1,2}^u, P_{3, \ldots, 6}$ interferences with $P_{7,9}$ via QED emission are infrared divergent. By the same argument, we discard these contributions, but keep the infrared finite $P_{1,2}^u, P_{3, \ldots, 6}$ interferences with $P_{8,10}$ via QED emission.
 

Finally, for $P_7$ and $P_8$ we use the scheme independent effective Wilson coefficients
\begin{align}\label{eq:C7C8}
C_7^{\rm eff}(\mu_b) {}& \equiv C_7(\mu_b) - \frac{1}{3} C_3(\mu_b) - \frac{4}{9} C_4(\mu_b) -\frac{20}{3}  C_5(\mu_b)- \frac{80}{9} C_6(\mu_b) \ ,
\nonumber \\ 
C_8^{\rm eff}(\mu_b) {}& \equiv C_8(\mu_b) + C_3(\mu_b) - \frac{1}{6} C_4(\mu_b) +20  C_5(\mu_b) - \frac{10}{3} C_6(\mu_b)  \, ,
\end{align}
which effectively also includes universal corrections from $b\bar{b}$ penguin loops. However, not taken into account are the finite, non-universal parts, induced for instance by the $c\bar{c}$ pair in the penguin loop, as well as those contributions where both the gluon and the photon are emitted from the penguin loop (see \cite{Huber:2014nna} for details). 

To summarize, we only consider the tree-level contributions of the $P_{1,2}^u, P_{3,\ldots, 6}$ operators and their interference with $P_{7, \ldots, 10}$. Infrared divergent parts that require fewer-particle cuts involving loops to render them finite are left for future study. We emphasize that this set of diagrams represents a gauge invariant subset.


\section{Phase-space integration}\label{sec:psint}

Having specified all contributions that we include in the present work, one has to calculate the squared matrix element of the process
\be
b(p_b) \, \to \, q(p_1) \, \bar q(p_2) \, d(p_3) \, \ell^-(p_4)  \, \ell^+(p_5)  \; , \label{eq:momenta}
\ee
and subsequently integrate it over the massless five-particle phase space~(PS). Due to the fact that we stay differential in the rescaled invariant-mass squared $q^2/m_b^2=(p_4+p_5)^2/m_b^2$ of the final-state lepton pair, the latter invariant acts as a regulator of infrared divergences in all interferences that we include. As a consequence, our calculation, and in particular the PS integration, can be done entirely in $D=4$ space-time dimensions. The completely differential five-particle PS measure reads
\be
d\Phi_5 \equiv \prod_{i=1}^5{d^3p_i\over(2\pi)^32E_i} (2\pi)^4\delta^4(p_b - \sum_{i=1}^5p_i) \, .
\ee

It turns out that our squared matrix element only depends on $m_b$, scalar products $(p_i \cdot p_j)$, $1\le i < j\le 5$, and linear factors $\epsilon_{\mu\nu\rho\sigma} p_i^\mu p_j^\nu p_k^\rho p_l^\sigma$. Moreover, the expression for $\cos\theta$ that we use to project onto the forward-backward asymmetry, can be written in terms of scalar products as well, see eq.~\Eqref{eq:costheta}. The fact that $d^3p_i$ contains all three-momentum configurations of particle $i$ ensures that, after PS integration, all terms that involve a {\emph{single}} $\epsilon$-tensor vanish, which also makes all functions of the final result being real-valued. Hence, from a practical point of view, we use PS parametrisations in terms of rescaled invariant masses $s_{ij} = 2p_i \cdot p_j/m_b^2 = (p_i+p_j)^2/m_b^2$, which are, by construction, dimensionless.

In our four-dimensional setup, there remain eight independent PS variables: The ten different $s_{ij}$, which are subject to the two constraints (a) that their sum equals to unity, which is a direct consequence of momentum conservation, and (b) that the Gram-determinant $\displaystyle \det\left[2(p_i\cdot p_j)\right]$, $i,j=1,\ldots,5$ vanishes. The latter condition holds since in $D=4$ dimensions, four light-like vectors are sufficient to span Minkowski space, and hence the five momenta $p_1^\mu,\ldots,p_5^\mu$ must be linearly dependent. In practice, we therefore have to perform seven integrations, keeping in mind that we stay differential in $s_{45}$. To this end, we adopt two different PS parametrisations, one by Kumar (K)~\cite{Kumar:1970cr}, and the other one by Heinrich (H)~\cite{Heinrich:2006sw}. We give details on both parametrisations in turn below.

\subsection{Integration according to Kumar}
\label{sec:PSK}
Following Kumar's parametrisation~\cite{Kumar:1970cr}, we have 
\begin{align}
\allowdisplaybreaks
\frac{d\Phi_5}{ds_3} = &\ {\pi^2m_b^6\over16(2\pi)^{11}}
\int_{s_3}^1 ds_1\int_{s_3}^{s_1} ds_2\int^{1-s_1+s_2}_{s_2/s_1}du_1\int^{u_{2}^+}_{u_{2}^-}du_2
\int^{u_{3}^+}_{u_{3}^-}du_3\int^{t_{2}^+}_{t_{2}^-}dt_2\int^{t_{3}^+}_{t_{3}^-}dt_3\nonumber\\
&\times \frac{(s_2-s_3)}{s_2 \, (u_2^+ - u_2^-) \, (u_3^+ - u_3^-) \, \sqrt{(t_2^+ - t_2) (t_2-t_2^-)} \; \sqrt{(t_3^+ - t_3) (t_3-t_3^-)}},
\end{align}
where the Lorentz-invariant Mandelstam-like variables are defined by 
\begin{align}
s_n &= {(p_b - \sum_{i=1}^np_i)^2\over m_b^2} \, , & u_n &= {(p_b - p_{n+1})^2\over m_b^2} \, , & t_n &= {(p_b - \sum_{i=2}^{n+1}p_i)^2\over m_b^2} \, , \label{eq:sijtosut} \\[0.2em]
s_2^\prime &= 1-{s_1}+{s_2}-{u_1} \, , & s_3^\prime &=2-{s_1}+{s_3}-{u_1}-{u_2} \, , & t_2^\prime &= 1+{t_2}-{u_1}-{u_2}  \, .
\end{align}
The integration limits are given by 
\begin{align}
\allowdisplaybreaks
   u_{2}^\pm = &\ 1-\frac{({s_1}+{u_1}) ({s_2}-{s_3})}{2{s_2}}\pm\frac{({s_2}-{s_3}) \sqrt{\lambda(1,{s_2},s_2^\prime)}}{2 {s_2}},\\
   u_{3}^\pm = &\ 1+\frac{1}{2} (1-{s_1}-{u_1}-{u_2})\pm\frac{1}{2} \sqrt{\lambda(1,{s_3},s_3^\prime)},\\
   t_{2}^\pm = &\ {u_1}-\frac{1}{2}({u_1}+1) (1-{u_2})+\frac{1}{2}(1-{u_1}) (1-{u_2})\left(\pm\sqrt{\left(1-{\eta_2^2}\right) \left(1-{\xi_2^2}\right)}-{\eta_2} {\xi_2}\right),\\
   t_{3}^\pm = &\ {t_2}-\frac{1}{2} (1-{u_3}) ({u_1}+{u_2}) + \frac{1}{2} (1-{u_3})\left(\pm\sqrt{\left(1-{\eta_3^2}\right) \left(1-\xi_3^2\right)}-{\eta_3} {\xi_3}\right) \!\sqrt{\lambda(1,{t_2},t_2^\prime)} \, .
\end{align}
Here
\be
\lambda(a,b,c) = a^2 + b^2 + c^2 - 2 a b - 2 a c - 2 b c
\ee
is the K\"all\'en function and
\begin{align}
\allowdisplaybreaks
\xi_2 & = \ \frac{(1-{u_1}) (2-{s_1}-{u_1})-2 s_2^\prime}{(1-{u_1}) \sqrt{\lambda(1,{s_2},s_2^\prime)}} \, , & \xi_3 = &\ \frac{(2-{u_1}-{u_2}) (1-{s_3}+s_3^\prime)-2(s_3^\prime+t_2^\prime)}{\sqrt{\lambda (1,{t_2},t_2^\prime)} \, \sqrt{\lambda(1,{s_3},s_3^\prime)}} \, , \nonumber\\
\eta_2 & =\ \frac{2 ({s_2}-{s_3})-(1-{u_2}) ({s_1}+{u_1})}{(1-{u_2})\sqrt{\lambda (1,{s_2},s_2^\prime)}} \, , & \eta_3 &= \ \frac{2 {s_3}-(1-{u_3})({s_1}+{u_1}+{u_2}-1)}{(1-{u_3}) \sqrt{\lambda (1,{s_3},s_3^\prime)}} \, .
\end{align}

Using eq.~\Eqref{eq:sijtosut}, we can now express the rescaled invariant masses $s_{ij}$ in terms of the integration variables $s_{1,2,3}$, $u_{1,2,3}$ and $t_{2,3}$. For all except one invariant, this is straightforward,
\begin{align}
s_{12} & =  {1-s_1+s_2-u_1} \, , & s_{13}  & = {u_1-s_2+s_3-t_2} \, , & s_{14} & = {t_2-s_3-t_3} \, , \nonumber \\
s_{24}  & = {1-{t_2}+{t_3}-{u_3}-s_{34}} \, , & s_{23} & = {1+t_2-u_1-u_2} \, , & s_{15} & = {t_3} \, , \nonumber \\
s_{25} & = {s_{34}+{s_1}-{s_2}-{t_3}+{u_1}+{u_2}+{u_3}-2} \, , & s_{35} & = -s_{34}+{s_2}-{s_3} \, , & s_{45} & = {s_3} \, . \label{eq:independent}
\end{align}
Equation~\Eqref{eq:independent} incorporates already the constraint that the sum of all ten $s_{ij}$ must equal to unity. The remaining invariant $s_{34}$ now gets extracted from the condition that the Gram determinant vanishes. Plugging eq.~\Eqref{eq:independent} into $\displaystyle \det\left[2(p_i\cdot p_j)\right] =0$ leaves us with a quadratic equation for $s_{34}$, whose two solutions read
\be
s_{34}^{\pm} = s_{34}^{r} \pm s_{34}^{s} \, ,
\ee
where $s_{34}^{r/s}$ refer to the rational and square-root part of the solution, respectively. The correct implementation of this two-fold solution in the PS integration then amounts to
\be
\displaystyle d\Phi_5 \, |\mathcal{M}|^2 \quad \to \quad d\Phi_5 \left(\frac{1}{2}|\mathcal{M}|^2_{\big|s_{34}\to s_{34}^{+}} + \frac{1}{2}|\mathcal{M}|^2_{\big|s_{34}\to s_{34}^{-}}\right) \, . \label{eq:s34kumar}
\ee

We note in passing that our results are at variance with some of the earlier works that dealt with the five-particle PS. The authors of~\cite{Achasov:2003re} obtained an expression for $s_{34}$ which captures only the $s_{34}^r$ part of the full solution. The reason can be traced back to the method proposed in Appendix D of~\cite{Kumar:1970cr}, where the delicate point is that only the {\emph{integration}}\footnote{We use the short-hand notation $p_{234}=p_2+p_3+p_4$ etc.\ here.}
\be
\int d^4p_4 \, \delta\!\left(p_4^2\right) \, \delta\!\left((p_b-p_{1234})^2\right) \, \delta\!\left((p_b-p_{234})^2-t_3\right) \, \delta\!\left((p_b-p_{4})^2-u_3\right) \, p_4^\mu
\ee
--~rather than $p_4^\mu$ {\emph{itself}}~-- is a linear combination of $p_b^\mu$, $(p_b-p_{123})^\mu$ and $(p_b-p_{23})^\mu$, although the conditions represented by the four delta functions in the integration truly have to be satisfied. Solving the four equations corresponding to the four delta functions, we find indeed two valid solutions for $p_4^\mu$, whose sum is a linear combination of $p_b^\mu$, $(p_b-p_{123})^\mu$ and $(p_b-p_{23})^\mu$. Therefore, what was obtained in \cite{Achasov:2003re} is actually the average of the two solutions for $s_{34}$.

\subsection{Integration according to Heinrich}
\label{sec:PSH}

The parametrisation of Heinrich~\cite{Heinrich:2006sw} assumes a form in which the PS measure completely factorises in the eight independent variables, which are labelled $t_2,\ldots,t_4,t_6,\ldots,t_{10}$ and whose integrations all run from $0$~to~$1$. In order to make the present paper self-contained, let us repeat the essential features from~\cite{Heinrich:2006sw} in the following.

The derivation starts from the PS parametrisation in $D$ dimensions, where only the constraint $\displaystyle s_{12} + \ldots + s_{45} =1$ holds, but not $\displaystyle \det\left[2(p_i\cdot p_j)\right] =0$. The relation between the rescaled invariant masses and the {\emph{nine}} variables $t_2,\ldots,t_{10}$ reads
\begin{align}
s_{13}&=t_6\,t_7\,(1-t_{2}) \, , & s_{15}&=t_7\,(1-t_6)\,[1-t_9\,(1-t_2t_4)]-y_{10} \, , \nonumber \\
s_{14}&=t_2\,t_4\,t_6\,t_7 \, , & s_{23}&=t_3\,(1-t_7)(1-t_2t_4)(t_6\,(1-t_9)+t_9) \, , \nonumber \\
s_{34}&=t_2\,t_6\,t_7\,(1-t_4) \, , & s_{25}&=y_8^-+(y_8^+-y_8^-)\,t_8 \, , \nonumber \\
s_{35}&=t_7\,t_9\,(1-t_6)(1-t_2t_4) \, , & s_{45}&=y_{10}^-+(y_{10}^+-y_{10}^-)\,t_{10}\, , \nonumber \\
s_{24}&=y_5^-+(y_5^+-y_5^-)\,t_5 \, , &  & \label{eq:sijti}
\end{align}
and $s_{12} = 1 - s_{13} - s_{14} - s_{15} - s_{23} - s_{24} - s_{25} - s_{34} - s_{35} - s_{45}$. The other abbreviations stand for
\begin{eqnarray}
y_8^\pm&=&y_8^0\pm d_8/2\nonumber\\
y_8^0&=&(1 - t_6)\,(1 - t_7)\,\{t_9 + t_3\,[t_6\,(1 - t_9) - t_9]\}/
     (t_6\,(1 - t_9) + t_9)\nonumber\\
d_8&=&y_8^+ -y_8^-=4\,(1-t_6)\,(1-t_7)\,
\sqrt{(1 - t_3)\,t_3\,t_6\,(1 - t_9)\,t_9}/
        (t_6\,(1-t_9) + t_9)\nonumber\\
y_{10}^\pm&=&y_{10}^0\pm d_{10}/2\nonumber\\
y_{10}^0&=&t_2\,t_7\,(1 - t_6)\,
\{1 - t_9 - t_4\,[1 -t_9 (2 - t_2)\,]\}/
      (1 - t_2\,t_4)\nonumber\\
d_{10}&=&y_{10}^+-y_{10}^-=4\,t_7\,t_2\,(1-t_6)\,
\sqrt{(1 - t_2)\,(1 - t_4)\,t_4\,
         (1 - t_9)\,t_9}/(1 - t_2\,t_4)\, ,
\end{eqnarray}
and $y_5^{\pm}$ are the solutions of $\displaystyle \det\left[2(p_i\cdot p_j)\right] =0$.

In the transition to $D=4$ dimensions the latter constraint is then implemented by letting $t_5$ assume the values $t_5=0$ or $t_5=1$ only, which renders $s_{24}=y_5^\pm$ and the number of independent variables is reduced to eight. The four-dimensional PS integral including a matrix element then reads
\begin{eqnarray}
\int d\Phi_5 \, |\mathcal{M}|^2&=&\frac{m_b^6}{4^8\pi^9}
\int\limits_0^1 dt_2 \ldots dt_4 dt_6 \ldots dt_{10}\,
\frac{t_2\,t_6\,t_7^2\,(1-t_6)(1-t_7)}{\sqrt{t_8} \, \sqrt{1-t_8} \, \sqrt{t_{10}} \, \sqrt{1-t_{10}}}\nonumber\\
&&\times\left(|\mathcal{M}|^2_{\big|t_5=0} + |\mathcal{M}|^2_{\big|t_5=1}\right) \, . \label{eq:t5heinrich}
\end{eqnarray}
Staying differential in one of the $s_{ij}$ is then simply taken into account by a corresponding delta-function factor, e.g.~$\delta(t_2\,t_4\,t_6\,t_7-s_{14})$, and hence the number of integrations that actually have to be performed is seven, as in the case of the parametrisation~(K).

\subsection{Comparison and concluding remarks}

Each of the two parametrisations has its virtues and its drawbacks. At a first glance, the parametrisation (H) seems superior to that of~(K): The integration limits are independent of each other and the symmetry under renaming of all final-state particles is manifest, which leaves us with a freedom to choose a labelling that makes the matrix element look as simple as possible.
In the parametrisation~(K), we use the freedom to rename $\{1,2,3\}$ (corresponding to $d$, $q$, $\bar q$) and $\{4,5\}$ (corresponding to the lepton pair). From a practical point of view, the parametrisation~(H) is easier when only massless propagators are present in the squared matrix element, whereas we prefer parametrisation~(K) whenever massive propagators occur. In appendix~\ref{sec:detailsPS} we illustrate with sample kernels how to perform the analytic integration in the~(H) and~(K) case, respectively.

Besides the analytic integration, we perform numerical checks in all cases. To this end we use a Monte-Carlo integration strategy with the Vegas~\cite{Lepage:1980dq} algorithm from the Cuba library~\cite{Hahn:2004fe}. We run the numerical integration for several values of $s_{45}$, typically with a million integration points. This gives us a relative precision of better than one percent, and in many cases even at the per-mill level. In the region of high $s_{45}$, the performance is a bit worse since the results rapidly approach zero due to the lack of PS volume.

\section{Results}
\label{sec:results}
In the following, we present our results for the branching ratio and the forward-backward asymmetry, both as functions of $\hat s \equiv q^2/m_b^2=(p_4+p_5)^2/m_b^2 = s_3  = s_{45}$. For the former observable, all results are obtained completely analytically in terms of transcendental functions of at most polylogarithmic weight three, while for the latter one numerical function required a fit.
\subsection{Branching ratio}
\label{sec:br}

The differential five-particle decay width can be expressed as 
\begin{align}
{d\Gamma(b\to d\ell^+\ell^-q\bar{q})\over d\hat{s}} 
= &\ {G_\text{F}^2m_{b,\text{pole}}^5\over48\pi^3}|V^*_{td}V_{tb}|^2\left(\sum_{i,j}\mathcal{R}^{ij}_\text{CKM}C_i^*C_j\mathcal{F}_{ij}(\hat{s})\right) \, .
\end{align}
Following \cite{Huber:2005ig}, we normalise the decay width to the branching ratio of the $\bar{B}\to X_ce\bar{\nu}$ decay, which 
is further expressed in terms of the perturbative expansion of the $\bar{B}\to X_ue\bar{\nu}$ decay (including power-corrections) and 
the ratio \cite{Gambino:2001ew,Bobeth:2003at}
\begin{align}
C = \left|{V_{ub}\over V_{cb}}\right|^2{\Gamma(\bar{B}\to X_ce\bar{\nu})\over \Gamma(\bar{B}\to X_ue\bar{\nu})}.
\end{align}
We use $C = 0.574\pm0.019$ \cite{Gambino:2013rza}.
Consequently, the expression of the branching ratio reads 
\begin{align}\label{eq:br}
{d\mathcal{B}(b\to d\ell^+\ell^-q\bar{q})\over d\hat{s}} 
= &\ \mathcal{B}(\bar{B}\to X_ce\bar{\nu})_\text{exp} \left|{V^*_{td}V_{tb}\over V_{cb}}\right|^2 {4\over C\Phi_u}
\left(\sum_{i,j=1}^{10}\mathcal{R}^{ij}_\text{CKM}C_i^*C_j\mathcal{F}_{ij}(\hat{s})\right) ,
\end{align}
where $\Phi_u$ is defined by \cite{Huber:2015sra, Huber:2005ig}
\begin{align}
\Gamma(\bar{B}\to X_ue\bar{\nu})
= &\ {G_\text{F}^2m_{b,\text{pole}}^5\over192\pi^3}|V_{ub}|^2\Phi_u.
\end{align}
This normalisation strategy is introduced 
to reduce the uncertainties arising from $m_{b,\text{pole}}$, CKM matrix elements, and phase-space factors involving $m_c$. We leave the details of the error analysis to a later work in which we perform a complete phenomenological study of the inclusive $\bar{B}\to X_d\ell^+\ell^-$ decay~\cite{inprep}.

The indices $i$ and $j$ in \eqref{eq:br} run over all the operators in eq.~\eqref{Leff}. For the Wilson coefficients $C_i(\mu)$, we take the values at the low 
scale $\mu = \mu_b = 5$~GeV. The analytical forms for $C_i(\mu_b)$ can be found in~\cite{Huber:2005ig}.
As explained in section~\ref{sec:me} we use $C_7^\text{eff}$ and $C_8^\text{eff}$ given by \eqref{eq:C7C8}. The CKM ratios are given as 
\begin{align}
\mathcal{R}^{ij}_\text{CKM} =\begin{cases}
               |\xi_u|^2, &\text{for}\ i, j = 1, 2;\\
               -\xi_u^*,&\text{for}\ i = 1, 2, j = 3, ..., 10;\\
               -\xi_u, &\text{for}\ i = 3, ..., 10, j = 1, 2;\\
               1, &\text{for}\ i, j = 3, ..., 10,\\
            \end{cases}
\end{align}
with $\xi_u \equiv  ({V_{ud}^* V_{ub}})/({V_{td}^*V_{tb}})$.
Finally, the tree-level contributions to the differential branching ratio from different operators are summarized in the 
$10\times 10$ matrix
\begin{align}
\allowdisplaybreaks
\mathcal{F}(\hat{s}) = &\left[
\begin{array}{ccccc}
\left(F_\text{cc}\right)_{2\times2} & \left(F_\text{cp}\right)_{2\times4} & \left(F_\text{cd}\right)_{2\times2} & \left(F_\text{cl}\right)_{2\times2} \\
\left(F_\text{pc}\right)_{4\times2} & \left(F_\text{pp}\right)_{4\times4} & \left(F_\text{pd}\right)_{4\times2} & \left(F_\text{pl}\right)_{4\times2} \\
\left(F_\text{dc}\right)_{2\times2} & \left(F_\text{dp}\right)_{2\times4} & \mathbf{0}_{2\times2} & \mathbf{0}_{2\times2} \\
\left(F_\text{lc}\right)_{2\times2} & \left(F_\text{lp}\right)_{2\times4} & \mathbf{0}_{2\times2} & \mathbf{0}_{2\times2} \\
\end{array}\right] \, ,
\end{align}
where
\begin{align}
\allowdisplaybreaks
&F_\text{cc} = \left[
\begin{array}{cc}
c_1U_1 & 0 \\
0 & c_2U_1 \\
\end{array}\right],\ \ 
F_\text{cp} = \left[
\begin{array}{cccc}
c_3U_1&  c_4U_1 & 16c_3U_1&  16c_4U_1 \\
U_1&  c_3U_1  & 16U_1&  16c_3U_1  \\
\end{array}\right], \ \
F_\text{pc} = \left[F_\text{cp}\right]^\text{T}, \nonumber \\
&F_\text{cd} = \left[
\begin{array}{cc}
c_4U_4 + c_3U'_1 & c_4U_5   \\
c_3U_4 + U'_1 & c_3U_5    \\
\end{array}\right],\ 
F_\text{dc} = \left[F_\text{cd}\right]^\text{T},\ 
F_\text{cl} = \left[
\begin{array}{cc}
 c_4U_6 + c_3U'_2 &  0  \\
 c_3U_6 + U'_2&  0   \\
\end{array}\right],\ 
F_\text{lc} = \left[F_\text{cl}\right]^\text{T},\nonumber \\
&F_\text{pp} = \left[
\begin{array}{cccc}
 c_2U_2 + U_3&  c_3U_3  & 10c_2U_2 + 16U_3 &  16c_3U_3  \\
c_3U_3 &  c_1U_2 +c_4U_3  & 16c_3U_3 &  10c_1U_2 +16c_4U_3  \\
10c_2U_2 + 16U_3 &  16c_3U_3  & 136c_2U_2 + 256U_3&  256c_3U_3  \\
16c_3U_3 &  10c_1U_2 +16c_4U_3  & 256c_3U_3 &  136c_1U_2 +256c_4U_3  \\
\end{array}\right], \nonumber \\
&F_\text{pd} = \left[
\begin{array}{cc}
c_2U'_6 + c_3U'_7 + U'_8 & c_3U'_3    \\
c_1U'_5 + c_4U'_7 + c_3U'_8 & c_1U_7 + c_4U'_3    \\
c_2U'_{10} + 16c_3U'_7 + 16U'_8 & 16c_3U'_3    \\
c_1U'_9 + 16c_4U'_7 + 16c_3U'_8 & c_1U'_4 + 16c_4U'_3    \\
\end{array}\right], \ \ 
F_\text{dp} = \left[F_\text{pd}\right]^\text{T}, \nonumber \\
&F_\text{pl} = \left[
\begin{array}{cc}
c_2U'_{12} + c_3U'_{13} + U'_{14} &  0   \\ c_1U'_{11} + c_4U'_{13} + c_3U'_{14} &  0   \\ c_2U'_{16} + 16c_3U'_{13} + 16U'_{14} &  0   \\ c_1U'_{15} + 16c_4U'_{13} + 16c_3U'_{14} &  0   \\
\end{array}\right],\ \ F_\text{lp} = \left[F_\text{pl}\right]^\text{T}.\ \  \label{matrix:br}
\end{align}
The explicit forms of all the $U^{(\prime)}_i(\s)$ functions can be found in Appendix \ref{sec:Ufunctions}. 
The color factors are defined by 
\begin{equation}
c_1 \equiv C_Ft_f, \qquad c_2 \equiv N_c, \qquad c_3 \equiv C_F, \qquad c_4 \equiv -{C_Ft_f\over N_c},
\end{equation}
with $N_c=3$, $t_f=1/2$, $C_F$ = 4/3. As discussed in section~\ref{sec:me}, the lower right $4\times4$ block, the last row and the last column of 
$\mathcal{F}(\hat{s})$ are absent.

For convenience, the $10\times10$ matrix $\mathcal{F}(\s)$ is provided in electronic form in the file {\tt FA.txt} that is attached to the arXiv submission of this article.


\subsection{Forward-backward asymmetry}
\label{sec:fba}

We define the forward-backward asymmetry $A_\text{FB}$ by 
\begin{align}
{dA_\text{FB}\over d\s} = &\ {1\over\Gamma^{\rm tot}_{\bar{B}}} \int_{-1}^1dz{d^2\Gamma(\bar{B}\to d\ell^+\ell^-q\bar{q})\over d\s dz}\text{sign}(z).
\end{align}
Here, $z = \cos\theta$, with $\theta$ the angle between the $\ell^+$ and $b$-quark three-momenta in the dilepton rest frame.
It turns out that $z$ can be expressed in terms of the Lorentz invariant momenta products as 
\begin{equation}
z = \cos\theta = \frac{s_{4b}-s_{5b}}{\lambda(1,s_{45},s_{123})^{1/2}}, \label{eq:costheta}
\end{equation}
with $s_{ib}\equiv 2p_i\cdot p_b/m_b^2$ and $s_{123} = (p_1+p_2+p_3)^2/m_b^2$. As anticipated in section~\ref{sec:psint} the structure of the squared
amplitude in all terms that we keep is such that projecting the double differential decay width onto the forward-backward asymmetry by means of the function $3/2 \, z$ is equivalent to
the projection using $\text{sign}(z)$, which we checked by explicit computation.

Using the same normalisation strategy as for the branching ratio \eqref{eq:br}, the forward-backward asymmetry is 
further expressed as 
\begin{align}\label{eq:afb}
{dA_\text{FB}\over d\hat{s}} = &\ \mathcal{B}(\bar{B}\to X_ce\bar{\nu})_\text{exp} \left|{V^*_{td}V_{tb}\over V_{cb}}\right|^2 {4\over C\Phi_u}
\sum_{i=1}^6\left( - \mathcal{R}_\text{CKM}^{i10}C_i^*C_{10} \mathcal{A}_i(\s) + c.c. \right),
\end{align}
with the tree-level contributions from different operators summarized in
\begin{align}
\allowdisplaybreaks
&\mathcal{A}(\s) = \mathcal{A}_\text{QCD}(\s) + \mathcal{A}_\text{QED}(\s)\, ,\ \ \label{matrix:afb}
\end{align}
with
\begin{align}
\allowdisplaybreaks
&\mathcal{A}_\text{QCD}(\s) = \left[
\begin{array}{cccccccc}
c_4U_8, & c_3U_8, & c_3U'_{18}, & c_1U'_{17} +c_4U'_{18}, & 16c_3U'_{18}, & 10c_1U'_{17} +16c_4U'_{18} \\
\end{array}\right]^\text{T}\, , \nonumber \\
&\mathcal{A}_\text{QED}(\s) = \left[
\begin{array}{cccccccc}
c_3U'_{19}, & U'_{19}, & c_2U'_{20} + U'_{21}, & c_3U'_{21}, & 10c_2U'_{20} + 16U'_{21}, & 16c_3U'_{21} \\
\end{array}\right]^\text{T}.
\end{align}
The functions $U^{(\prime)}_i(\s)$ are again relegated to Appendix~\ref{sec:Ufunctions}, and the quantity $\mathcal{A}(\s)$ is also provided in electronic form in the file {\tt FA.txt}.


\section{Numerical estimate and conclusion }
\label{sec:conclusion}

In this section we give the numerical estimates for the branching ratio and the forward-backward asymmetry 
of the five-particle decay process $b\to d\ell^+\ell^-q\bar{q}$. For each observable we give the integral over bin~1 
($q^2 \in [1,3.5]$~GeV$^2$), bin~2 ($q^2 \in [3.5,6]$~GeV$^2$), and the entire 
low-$q^2$ region ($q^2 \in [1,6]$~GeV$^2$). We do not consider the high-dilepton invariant-mass region here since there the five-body contributions are negligible because they are suppressed by high powers of $(1-\hat s)$ due to the lack of PS volume. We use the input parameters from Table 1 of 
\cite{Huber:2015sra}, except that for the CKM matrix elements we take the most recent determination of the parameters \cite{Koppenburg:2017mad}
\begin{align}
\lambda = 0.2251^{+0.007}_{-0.012},\ A = 0.825\pm 0.0003,\ \bar{\rho} = 0.160^{+ 0.008}_{- 0.007},\ \bar{\eta}= 0.350\pm0.006,
\end{align}
which gives
\begin{equation}
\left|{V^*_{td}V_{tb}\over V_{cb}}\right|=0.204719,\qquad \xi_u   =  0.0143702 -0.422654i.
\end{equation}
Then, we obtain the numerical results for the branching ratio and the forward-backward asymmetry in the three bins, as shown in Table \ref{tab:numrst}. Only the central values are presented, whereas the study of uncertainties is relegated to a forthcoming work focusing on the phenomenological analysis of the inclusive $\bar{B}\to X_d\ell^+\ell^-$ decay~\cite{inprep}.

\begin{table}[t] 
\begin{center}
\begin{tabular}{cccc} \hline \hline
 & [1, 3.5] GeV$^2$  &   [3.5, 6] GeV$^2$  & [1, 6] GeV$^2$   \\ \hline  
$\mathcal{B}(b\to d\ell^+\ell^-q\bar{q})$ ($\times10^{-10}$)    & 9.22    & 0.30 & 9.52  \\  
$A_\text{FB}(b\to d\ell^+\ell^-q\bar{q})$ ($\times10^{-12}$)     & 1.48 & 0.49  & 1.97  \\  
\hline \hline
\end{tabular}
\caption{Numerical estimates for the branching ratio and the forward-backward asymmetry of $b\to d\ell^+\ell^-q\bar{q}$ in the low-$q^2$ region.}\label{tab:numrst} 
\end{center}
\end{table}

Keeping only the $P_{1,2}^u, P_{3,\ldots,6}$ interferences among themselves, which start contributing at order $\mathcal{O}(\widetilde\alpha^2_s \kappa^2)$, we find for the low-$q^2$ integrated branching ratio $\mathcal{B}(b\to d\ell^+\ell^-q\bar{q})_{[1,6]} = 9.60\times10^{-10}$. Actually, if we furthermore drop the penguin operators $P_{3,\ldots,6}$ we already obtain $9.40\times10^{-10}$ in the same $q^2$ region, which demonstrates that the $P^u_{1,2}$~--~$P^u_{1,2}$ interferences dominate.

Comparing the five-particle branching ratio and the three-particle branching ratio of $\bar{B}\to X_d\ell^+\ell^-$ that we have also 
estimated, we find that $\mathcal{B}(b\to d\ell^+\ell^-q\bar{q})_{[1,6]}$ contributes about 1.4\% to 
$\mathcal{B}(\bar{B}\to X_d\ell^+\ell^-)_{[1,6]}$ and $\mathcal{B}(b\to d\ell^+\ell^-q\bar{q})_{[1,3.5]}$ contributes about 2.5\% to 
$\mathcal{B}(\bar{B}\to X_d\ell^+\ell^-)_{[1,3.5]}$.

Our results are also valid for $b\to s\ell^+\ell^-q\bar{q}$, and can be obtained by changing the CKM matrix elements $V_{ud}\to V_{us}$ and $V_{td}\to V_{ts}$ in Eqs.~\eqref{eq:br} and \eqref{eq:afb}. The ratios relevant for the calculation are
\begin{equation}
\left|{V^*_{ts}V_{tb}\over V_{cb}}\right|= 0.981942,\qquad \frac{{V_{us}^* V_{ub}}}{{V_{ts}^*V_{tb}}}  = -0.00879872 + 0.0183709i \, .
\end{equation}
We straightforwardly obtain the corresponding results for $b\to s\ell^+\ell^-q\bar{q}$, as listed in Table \ref{tab:numrst2}. 
Compared to the lastest theory prediction for the branching ratio in the low-$q^2$ region~\cite{Huber:2015sra}, 
\begin{align}
\mathcal{B}(\bar{B}\to X_s\ell^+\ell^-)_{[1,6]} = (1.62 \pm 0.09)\times 10^{-6},
\end{align}
the five-particle contribution is at the level of $\mathcal{O}(0.01\%)$, which shows the expected CKM suppression.

To conclude, we have presented the first study of tree-level five-particle contributions to $\bar{B}\to X_{s(d)} \ell^+\ell^-$. While for the $b\to s$ transition, such contributions are numerically at the sub-permille level, they play a more pronounced role in $b\to d$ where their contribution is estimated to be at the percent level.
In light of the anomalies in exclusive $b \to s \ell\ell$ systems and the upcoming data-taking at Belle~II, a study of inclusive $\bar{B}\to X_{s(d)} \ell^+\ell^-$
decays is both timely and relevant since it provides important complementary constraints compared to the widely studied exclusive decays.
The present study paves the road for a phenomenological update of both the branching ratio and the forward-backward asymmetry in $\bar{B}\to X_{d} \ell^+\ell^-$~\cite{inprep}, whose latest analysis dates back 15 years~\cite{Asatrian:2003vq}.

From a computational point of view, the integrations of the squared matrix elements over the massless five-particle phase-space in four dimensions, while simultaneously staying differential in one of the kinematic invariants, is very challenging. Our paper therefore serves as a proof-of-concept that such integrations can --~with one exception~-- be carried out completely analytically in terms of polylogarithmic functions of at most weight three.

\begin{table}[t] 
\begin{center}
\begin{tabular}{cccc} \hline \hline
 & [1, 3.5] GeV$^2$  &   [3.5, 6] GeV$^2$  & [1, 6] GeV$^2$   \\ \hline  
$\mathcal{B}(b\to s\ell^+\ell^-q\bar{q})$ ($\times10^{-10}$)    & 2.18    & 0.05 & 2.23  \\  
$A_\text{FB}(b\to s\ell^+\ell^-q\bar{q})$ ($\times10^{-11}$)     & 1.57  & 0.52 & 2.10  \\  
\hline \hline
\end{tabular}
\caption{Numerical estimates for the branching ratio and the forward-backward asymmetry of $b\to s\ell^+\ell^-q\bar{q}$ in the low-$q^2$ region.}\label{tab:numrst2} 
\end{center}
\end{table}


\subsubsection*{Acknowledgements}

We would like to thank Thomas Gehrmann, Gudrun Heinrich, Tobias Hurth, Matthias Jamin, Enrico Lunghi and Javier Virto for useful discussions. This work is supported by the Deutsche Forschungsgemeinschaft (DFG) within research unit FOR 1873 (QFET).



\begin{appendix}

\section{Details on the phase-space integration}
\label{sec:detailsPS}

In this appendix we describe how to integrate two sample kernels analytically over the five-particle PS, using the parametrisations (K) and (H), respectively.

\subsection{Integration via parametrisation (K)}
\label{sec:detailskumar}

As mentioned in the main text, the parametrisation (K) turns out to be quite efficient if massive propagators are present. We therefore illustrate the integration $\displaystyle \frac{d\Phi_5}{ds_3} K$ of the kernel
\begin{equation}
K = \frac{(4\pi)^7}{m_b^6} \, \frac{s_{34}}{(s_{12}+s_{13}+s_{23}-1) \, (s_{34}+s_{35}+s_{45})} \, ,
\end{equation}
which appears for instance in the $P_2^u$~--~$P_2^u$ interference. We substitute the rescaled invariant masses $s_{ij} = (p_i+p_j)^2/m_b^2$ according to section~\ref{sec:PSK}, and apply in particular eq.~(\ref{eq:s34kumar}) for the treatment of $s_{34}$.

First, we perform the $t_3$-integration, where it turns out that the dependence of $K$ on $t_3$ is polynomial. After the substitution
\begin{equation}
t_3 = (t_3^+-t_3^-) \chi + t_3^- \, ,
\end{equation}
the $\chi$-integration runs from $0$ to $1$ and can be carried out in terms of $\Gamma$-functions, yielding
\begin{equation}
\int^{t_{3}^+}_{t_{3}^-}dt_3  \frac{1}{\sqrt{(t_3^+ - t_3) (t_3-t_3^-)}} = \pi \, ,
\end{equation}
and similar if there is a polynomial in $t_3$ in the numerator.

The obtained integrand has a polynomial dependence on $u_3$. Since the limits $t_2^\pm$ of the $t_2$-integration are independent of $u_3$, we can trivially interchange the two integrations, and the $u_3$-integration is elementary (i.e.\ one computes an integral function and plugs in the limits). Surprisingly, after these two integrations, the integrand has simply become
\begin{equation}
-\frac{s_2-s_3}{4 \pi  s_2 \, (s_1-s_3+u_1+u_2-1) \, \sqrt{(s_1+u_1)^2-4 s_2} \, \sqrt{(t_2^+ - t_2) (t_2-t_2^-)}} \, .
\end{equation}
Looking at its structure, we see immediately that we can perform the $t_2$-integration along the same lines as the $t_3$-integration before. However, we experienced also cases where the dependence on $t_2$ at this stage is more complicated. For instance, a $1/t_2$-dependence can occur, which makes the subsequent substitutions considerably more involved.

The next integration to be performed is that over $u_2$, which is also elementary, but which now introduces a logarithm. We observe, however, that the dependence of the resulting integrand on the variables $s_1$ and $u_1$ is via the combination $u_1+s_1$ only! We therefore shift the variable $u_1 \to \widetilde u_1 - s_1$ and interchange the order of the remaining integrations, which results in the new integration limits $s_2 = s_3 \ldots 1$, $\widetilde u_1 = 2\sqrt{s_2} \ldots 1+s_2$, and $s_1^\pm = \widetilde u_1/2 \pm \sqrt{\widetilde u_1^2-4 s_2}/2$. The integration over $s_1$ is then trivial, yielding the integrand
\begin{equation}
\frac{(s_2-s_3)}{4 s_2} \, \ln \left(\frac{\widetilde u_1(s_2+s_3)-2 s_2 s_3 - (s_2-s_3) \sqrt{\widetilde u_1^2-4 s_2}}
   {\widetilde u_1(s_2+s_3)-2 s_2 s_3 + (s_2-s_3) \sqrt{\widetilde u_1^2-4 s_2}}\right) \, .
\end{equation}

The remaining two integrations over $\widetilde u_1$ and $s_2$, which we perform in this order, are also elementary if a computer algebra system is used. After simplification, the final result reads
\begin{align}
\frac{d\Phi_5}{ds_3} K  &= \frac{3}{8} \sqrt{(4-s_3) s_3} \, (1-s_3\ln(s_3)) f_1(s_3) +\frac{3}{8} (s_3-2) s_3
   [f_1(s_3)]^2-\frac{1}{16} (s_3-1) (5 s_3-1) \nonumber \\[0.3em]
   &-\frac{1}{32} s_3 (3 s_3-2) \ln^2(s_3)+\frac{7}{16} s_3 \ln(s_3)\, ,
\end{align}
where $f_1$ is defined in eq.~\Eqref{eq:f1}.

\subsection{Integration via parametrisation (H)}
\label{sec:detailsheinrich}

The parametrisation (H) is efficient in the case of only massless propagators. We therefore demonstrate how to integrate the sample kernel
\begin{equation}
H = \frac{(4\pi)^7}{m_b^6} \, \frac{s_{35}}{(s_{14}+s_{15}+s_{45}) \, (s_{24}+s_{25}+s_{45})} \, ,
\end{equation}
which also appears in the $P_2^u$~--~$P_2^u$ interference. To this end, we first relabel the particle indices in the PS parametrisation eq.~\Eqref{eq:sijti} (but {\emph{not}} in $H$) according to $\{1,2,3,4,5\} \to \{5, 3, 2, 4, 1\}$. Staying differential in the lepton invariant-mass squared is then implemented via the factor $\delta(t_2\,t_4\,t_6\,t_7-s_{45})$, and implementing the vanishing of the Gram determinant proceeds via eq.~\Eqref{eq:t5heinrich}.

The first two integrations are over $t_8$ and $t_{10}$. In both cases, the dependence of the integrand on these variables is of the type $t_i^a (1-t_i)^b$, with integer or half-integer $a$ and $b$. Hence, both integrations can be done in terms of $\Gamma$-functions. Afterwards, the integrand is polynomial in $t_3$ and integration results in ($\bar t_i = 1-t_i$)
\begin{equation}
\left[\frac{t_2 \bar t_6 \bar t_7^{\, 2} \left(\bar t_2-t_2^2 t_4 \bar t_4\right)}{4 (1-t_2 t_4)^2}
   -\frac{t_2 \bar t_6 \bar t_7^{\, 2} \left(\bar t_2 (t_2 t_4 (t_6+1)-t_6)-t_2^2 t_4 \bar t_4\right)}{4 (1-t_2 t_4)^2 (1-(1-t_2 t_4) (t_6 \bar t_9+t_9))} \right] \, \delta(t_2\,t_4\,t_6\,t_7-s_{45})\, .
\end{equation}

The next integration over $t_9$ is elementary and introduces a logarithm. We observe that in the resulting integrand, the dependence on the variables $t_2$ and $t_4$ in the denominator and in the $\delta$-function is via the combination $t_2t_4$ only! We therefore substitute $t_2 = w/t_4$ and integrate over $t_4$ next, i.e.
\begin{equation}
\int\limits_0^1\!\!dt_4\int\limits_0^1\!\!dt_2 = \int\limits_0^1\!\!dw \int\limits_w^1\!\!dt_4 \, \frac{1}{t_4}\, .
\end{equation}
The resulting expression is short,
\begin{equation}
\left[\frac{1}{8} \bar t_6 \bar t_7^{\, 2} \bar w-\frac{1}{8} t_6 \bar t_7^{\, 2} \ln
   \left(\frac{w}{1-t_6 \bar w}\right) \right] \, \delta(w\,t_6\,t_7-s_{45})\, .
\end{equation}

At this stage, we apply the above trick again twice by first substituting $t_7 = x/t_6$ (followed by integration over $t_6$ from $x \ldots 1$) and subsequently setting $w = z/x$, followed by integrating over $x$ from $z \ldots 1$. Both integrations are elementary with a computer algebra system at hand. What remains is a trivial integration over $z$ due to the factor $\delta(z-s_{45})$. After simplification, the final result reads
\begin{align}
\frac{d\Phi_5}{ds_{45}} H  &= \frac{1}{16} (s_{45}-31) (s_{45}-1)+\frac{1}{8} (1-2 s_{45}) \ln ^2(s_{45})+\frac{1}{8} (8 s_{45}+7) \ln(s_{45})\, .
\end{align}
\section{Functions}
\label{sec:Ufunctions}

Here we present the $U^{(\prime)}$ functions appearing in \eqref{matrix:br} and \eqref{matrix:afb}. The electric charge factors are $Q_L=-1$, $Q_d=Q_s=-1/3$, $Q_u=+2/3$, and the sum over $q$ runs over $u,d,s$.
\begin{align}
\allowdisplaybreaks
U_1 = &\ \tilde{\alpha}_s^2\kappa^2 Q_L^2 \bigg[ \frac{Q_d Q_u}{81}  \bigg(-108 \sqrt{4-\s} \sqrt{\s} \left(4 \s^2-7 \s-6\right)
   f_1(\s) \ln (\s)
   \nonumber \\
   & +216 \sqrt{4-\s} \sqrt{\s} (4 \s-1) f_1(\s)-72 (\s+1) \left(4 \s^2+5
   \s-5\right) f_2(\s)+432 f_3(\s) \nonumber \\
   & -\frac{(\s-1) \left(955 \s^3-425
   \s^2-749 \s+87\right)}{\s}-9 \left(12 \s^3-45 \s^2-10\right) \ln ^2(\s) \nonumber \\
   & +\frac{12 \left(84 \s^3-15 \s^2-56 \s+3\right) \ln (\s)}{\s} +108 \left(4 \s^3-15 \s^2+10\right) [f_1(\s)]^2 \bigg) \nonumber \\
   & +\frac{Q_d^2}{324} \bigg(216 \sqrt{4-\s} \sqrt{\s} \left(\s^2+2 \s-12\right) f_1(\s) \ln
   (\s)-\frac{216 \left(\s^4-18 \s^2+16 \s+6\right) [f_1(\s)]^2}{\s} \nonumber \\
   & +\frac{108 \left(4 \s^3+13 \s^2-110 \s-60\right) f_1(\s)}{\sqrt{4-\s}
   \sqrt{\s}}+\frac{18 \left(3 \s^4-18 \s^2+16 \s+6\right) \ln^2(\s)}{\s} \nonumber \\
   & +\frac{(\s-1) \left(259 \s^3+43 \s^2+133 \s-1863\right)}{\s}
   -\frac{6 \left(36 \s^3+477 \s^2-172 \s-120\right) \ln(\s)}{\s}\bigg)\nonumber \\
   & +\frac{Q_u^2}{324}  \bigg(\frac{144 (\s+1) \left(\s^3+5 \s^2+13
   \s-3\right) f_2(\s)}{\s}+\frac{72 \left(18 \s^2+10 \s-3\right) \ln ^2(\s)}{\s}\nonumber \\
   &+1728 f_3(\s)+\frac{(\s-1) \left(259 \s^3+715 \s^2+6835\s-2901\right)}{\s}\nonumber \\
   &-\frac{12 \left(24 \s^3+348 \s^2+148 \s-111\right) \ln(\s)}{\s}\bigg) \bigg] \, , \\[1.0em]
U_2 = &\ \tilde{\alpha}_s^2\kappa^2 Q_L^2 \sum_{q}\bigg[ \frac{Q_d^2}{162} \bigg(
216 \sqrt{4-\s} \sqrt{\s} \left(\s^2+2 \s-12\right)f_1(\s) \ln (\s)
   \nonumber \\
   & -\frac{216 \left(\s^4-18 \s^2+16 \s+6\right)[f_1(\s)]^2}{\s} -\frac{6 \left(36 \s^3+477 \s^2-172 \s-120\right) \ln(\s)}{\s} \nonumber \\
   & +\frac{108 \left(4 \s^3+13 \s^2-110 \s-60\right)
   f_1(\s)}{\sqrt{4-\s} \sqrt{\s}}+\frac{18 \left(3 \s^4-18 \s^2+16 \s+6\right)\ln ^2(\s)}{\s}\nonumber \\
   & +\frac{(\s-1) \left(259 \s^3+43 \s^2+133\s-1863\right)}{\s}
   \bigg)\nonumber \\
   & +\frac{Q_q^2}{162} \bigg(\frac{144 (\s+1) \left(\s^3+5 \s^2+13
   \s-3\right) f_2(\s)}{\s}+\frac{72 \left(18 \s^2+10\s-3\right) \ln ^2(\s)}{\s}\nonumber \\
   & +1728 f_3(\s)+\frac{(\s-1) \left(259 \s^3+715 \s^2+6835\s-2901\right)}{\s}\nonumber \\
   & -\frac{12 \left(24 \s^3+348 \s^2+148 \s-111\right) \ln(\s)}{\s}\bigg) \bigg] \, , \\[1.0em]
U_3 = &\ \tilde{\alpha}_s^2\kappa^2 Q_L^2 \frac{Q_d^2}{162}  \bigg(-108 \sqrt{4-\s} \sqrt{\s} \left(7 \s^2-16 \s\right)f_1(\s) \ln (\s)
   +1728 f_3(\s) \nonumber \\
   & -\frac{72 (\s+1) \left(7 \s^3+5 \s^2-23\s+3\right) f_2(\s)}{\s}
   -\frac{54 \left(28 \s^3-149 \s^2+142 \s+60\right)f_1(\s)}{\sqrt{4-\s} \sqrt{\s}}  \nonumber \\
   &+\frac{3 \left(588 \s^3-1293 \s^2-572 \s+366\right) \ln (\s)}{\s}
   -\frac{9 \left(21 \s^4-90 \s^3-54 \s^2-76\s+6\right) \ln ^2(\s)}{\s} \nonumber \\
   &+\frac{108 \left(7 \s^4-30 \s^3+18 \s^2+4\s-6\right) [f_1(\s)]^2}{\s} \nonumber \\
   &-\frac{(\s-1) \left(1651\s^3-1229 \s^2-4982 \s+2556\right)}{\s}
   \bigg)\, ,\\ 
U_4 = &\ \tilde{\alpha}_s^3\kappa^2Q_L^2\bigg[\frac{2 Q_d}{27 \s^{3/2}}  \left(-108 \sqrt{\s} \left(\s^3-2 \s^2-5 \s+12\right)f_1(\s)^2 
\right. \nonumber\\ 
&\left. +18  (\s-1)^2 (\s+2) \sqrt{4-\s}f_6(\s)+2 (\s-1) \left(37 \s^2-359 \s+64\right) \sqrt{\s}\right. \nonumber\\
& \left. +12 \left(47 \s^2-19 \s+6\right) \sqrt{\s} \log (\s)+9 \left(\s^3-10 \s^2+23 \s-12\right) \sqrt{\s} \log ^2(\s) \right. \nonumber\\
& \left. +9 (\s-1)^2 (\s+12) \sqrt{\s}f_4(\s)  
+216(\s-2) \s \sqrt{4-\s}f_1(\s) 
\right)\nonumber\\
& +\frac{1}{3} Q_u \bigg(-96\s f_5(\s)+6 (\s-1) (9 \s-1) f_4(\s) +\frac{(\s-1) \left(179 \s^2+227 \s+8\right)}{\s} \nonumber \\
&  +6 (2 \s-1) \log ^2(\s)-6 (56 \s+13) \log (\s)\bigg)	\bigg],  \\
U_5 = &\ \tilde{\alpha}_s^3\kappa^2Q_L^2\bigg[
Q_dQ_u \left(12 ({\s}-2) (2 {\s}-3) {f_1}({\s})^2-32\sqrt{4-{\s}} \sqrt{{\s}} {f_1}({\s})+16 ({\s}-1) {f_3}({\s})\right. \nonumber\\
   &-12\sqrt{4-{\s}} \sqrt{{\s}} (2 {\s}-3) {f_1}({\s}) \log ({\s})+8 ({\s}+1) (2 {\s}-3){f_2}({\s}) -\frac{2}{3} ({\s}-1) \log ^3({\s}) \nonumber\\
   &+24 ({\s}-1) {f_7}({\s})+\frac{2 ({\s}-1) \left(3{\s}^2-175 {\s}+10\right)}{3 {\s}}+\left(-6 {\s}^2+13 {\s}-10\right) \log^2({\s})\nonumber\\
   &\left.+\frac{4 \left(34{\s}^2+37 {\s}-2\right) \log ({\s})}{3 {\s}} -24 ({\s}-1) {f_1}({\s})^2 \log ({\s}) \right) \nonumber \\
  & +Q_d^2  \left(-\frac{4 \left(25 {\s}^2-94 {\s}-60\right) {f_1}({\s})}{3\sqrt{4-{\s}} \sqrt{{\s}}}+\frac{8 \left({\s}^3-4 {\s}^2+2 {\s}+2\right){f_1}({\s})^2}{{\s}} \right. \nonumber\\
   &-8 \sqrt{4-{\s}} ({\s}-2) \sqrt{{\s}} {f_1}({\s}) \log({\s})-\frac{2 ({\s}-1) \left(122 {\s}^2+89 {\s}-295\right)}{27 {\s}} \nonumber\\
   &\left.+\frac{2 \left(155 {\s}^2-16{\s}-38\right) \log ({\s})}{9 {\s}}-\frac{2 \left(3 {\s}^3-4 {\s}^2+2 {\s}+2\right) \log^2({\s})}{3 {\s}}\right) \nonumber\\
&+ Q_u^2 \left( \frac{16 ({\s}+1)\left({\s}^2-10 {\s}+1\right){f_2}({\s})}{3 {\s}}  +\frac{4({\s}-1) \left(31 {\s}^2-1391 {\s}+274\right)}{27 {\s}} \right. \nonumber\\
&\left.+32 ({\s}-1) {f_3}({\s}) -\frac{8 \left(9 {\s}^2+9 {\s}-1\right) \log^2({\s})}{3 {\s}} 
   +\frac{8 \left(114 {\s}^2+87 {\s}-20\right) \log ({\s})}{9 {\s}}\right)	\bigg], \\
U_6 = &\ \tilde{\alpha}_s^2\kappa Q_L\bigg[
Q_d\left( \frac{4 \left(2 {\s}^4-3 {\s}^3-12 {\s}^2+15 {\s}+12\right){f_1}({\s})^2}{3 {\s}} - \frac{8 \sqrt{4-{\s}} ({\s}-2) ({\s}+1) {f_1}({\s})}{3\sqrt{{\s}}} \right. \nonumber\\
&-\frac{({\s}+12) (2 {\s}+1) ({\s}-1)^2 {f_4}({\s})}{9 {\s}} - \frac{2\sqrt{4-{\s}} ({\s}+2) (2 {\s}+1) ({\s}-1)^2 {f_6}({\s})}{9 {\s}^{3/2}} \nonumber\\
&-\frac{2}{27} \left(72{\s}^2-42 {\s}-43\right) \log ({\s}) - \frac{\left(62 {\s}^3-613 {\s}^2+269 {\s}+360\right)({\s}-1)}{81 {\s}} \nonumber\\
&-\frac{\left(62 {\s}^3+17 {\s}^2-253 {\s}-216\right) ({\s}-1)}{81{\s}}-\frac{2 \left(24 {\s}^3-12 {\s}^2+5 {\s}+12\right) \log ({\s})}{27 {\s}} \nonumber\\
&\left.-\frac{\left(2{\s}^4-3 {\s}^3+36 {\s}^2-{\s}-12\right) \log ^2({\s})}{9 {\s}}\right) \nonumber\\
&   +Q_u \left(-({\s}-1) \left(8 {\s}^2+{\s}-1\right) {f_4}({\s})-\frac{1}{18} ({\s}-1) \left(484 {\s}^2+457 {\s}+61\right) \right.\nonumber\\
&\left.+16 {\s}^2{f_5}({\s})+\frac{1}{3} \left(132 {\s}^2+36{\s}-1\right) \log ({\s})-(2 {\s}-1) \log ^2({\s})\right) \bigg], \\
U_7 \equiv &\ \sum_q \hat{U}_7 =\tilde{\alpha}_s^3\kappa^2Q_L^2\sum_q\bigg[
Q_d^2 \bigg( 
\frac{8 \left(-25 {\s}^2+94 {\s}+60\right) {f_1}({\s})}{3\sqrt{4-{\s}} \sqrt{{\s}}} 
+\frac{4 \left(155 {\s}^2-16{\s}-38\right) }{9 {\s}}  \nonumber\\
& \times\log ({\s})-16 \sqrt{4-{\s}} ({\s}-2) \sqrt{{\s}} {f_1}({\s}) \log({\s}) 
- \frac{4 ({\s}-1) \left(122 {\s}^2+89 {\s}-295\right)}{27 {\s}}\nonumber\\
&+ \frac{16 \left({\s}^3-4 {\s}^2+2 {\s}+2\right){f_1}({\s})^2}{{\s}}
-\frac{4 \left(3 {\s}^3-4 {\s}^2+2 {\s}+2\right) \log^2({\s})}{3 {\s}}\bigg)\nonumber\\
  & +Q_q^2 \bigg(\frac{32 ({\s}+1)\left({\s}^2-10 {\s}+1\right) {f_2}({\s})}{3 {\s}}   
-\frac{16 \left(9 {\s}^2+9 {\s}-1\right) \log^2({\s})}{3 {\s}}
  +64 ({\s}-1) {f_3}({\s}) \nonumber\\
& +\frac{8({\s}-1) \left(31 {\s}^2-1391 {\s}+274\right)}{27 {\s}}
   +\frac{16 \left(114 {\s}^2+87 {\s}-20\right) \log ({\s})}{9 {\s}} \bigg)\bigg],  \\
U_8=&\ \tilde{\alpha}_s^2\kappa {Q_dQ_L\over 72} \bigg[
   12 \left(2 \s^2-6 \s+5\right) \log(2-\sqrt{\s})-\frac{2 \left(2 \s^4+8 \s^3+24 \s^2+5\s+4\right) f_8(\s)}{\s} \nonumber \\
   &-\frac{3 (\s-4) (\s-1)^2 (2 \s+1)f_9(\s)}{\s} +\frac{6x \left(2 \s^4-15 \s^3+30 \s^2+3\s-18\right) f_{10}(\s)}{\s} \nonumber \\
   &-\frac{\left(48 \s^{5/2}-187 \s^{3/2}+114 \s-95 \sqrt{\s}+12\right)\left(\sqrt{\s}-1\right)^2}{\sqrt{\s}}+6 (\s-8) \s^2 \log^2(\s) \nonumber \\
   &+2 \left(28 \s^{5/2}+12 \s^{3/2}-52 \s-12(\s-1)^2 \s \log \left(\sqrt{\s}+1\right)+15\right) \log (\s) \nonumber \\
   &-\frac{4\left(4 \s^{5/2}+22 \s^{3/2}+72 \s^2+57 \s-26\sqrt{\s}+3\right) \left(\s-1\right) \log \left(\sqrt{\s}+1\right)}{\s}\bigg] \nonumber \\
   &+\tilde{\alpha}_s^2\kappa {Q_uQ_L} \bigg[ (\s-1)\left(127.78 \s^3-19268. \s^2-99.381 \s-0.39678\right) \nonumber \\
   &+\sqrt{\s} \left(3.1004 \s^3+483.32 \s^2+21.715 \s-0.0011409\right) \log^3(\s) \nonumber \\
   &+\sqrt{\s} \left(-137.22 \s^3+51.514 \s^2+188.38 \s-0.037079\right) \log ^2(\s) \nonumber \\
   &+\sqrt{\s} \left(440.67 \s^3+18028. \s^2+771.90\s-0.31575\right) \log (\s) \bigg] \, .
\end{align}
The term proportional to $Q_u$ in $U_8$ was obtained from a least-square fit.
\begin{align}
U'_1  &=\kappa \left(Q_dQ_uU_4 + Q_uU_5\right),\  
&U'_2  &= \kappa Q_dQ_uU_6, \   \nonumber \\
U'_3  &= U_5|_{Q_u\to Q_d},  
&U'_4 &=  4U_7 + 12 \sum_{q}U_5|_{Q_u\to Q_q}, \ \nonumber \\
U'_5 &\equiv \sum_q \hat{U}'_5=  2\sum_{q}U_4|_{Q_u\to 0}, \ 
&U'_6 &= \sum_q \kappa\left(Q_dQ_q\hat{U}'_5 + Q_q\hat{U}_7\right),  \nonumber \\
U'_7 &= U_4|_{Q_u\to Q_d}, \ 
&U'_8 &= \kappa\left(Q_d^2U'_7 + Q_dU'_3\right),\  \nonumber \\
U'_9 &= 4U'_5 + 12 \sum_{q}U_4|_{Q_u\to Q_q},  
&U'_{10} &= 4U'_6 + 12 \sum_{q}U'_1|_{Q_u\to Q_q}, \  \nonumber \\
U'_{11} &\equiv \hat{U}'_{11}= 2\sum_{q}U_6|_{Q_u\to 0}, \ 
&U'_{12} &= \sum_q\kappa Q_dQ_q\hat{U}'_{11},  \nonumber \\
U'_{13} &= U_6|_{Q_u\to Q_d}, \ 
&U'_{14} &= U'_2|_{Q_u\to Q_d} = \kappa Q_d^2U'_{13}, \  \nonumber \\
U'_{15} &= 4U'_{11} + 12 \sum_{q}U_6|_{Q_u\to Q_q} , \
&U'_{16} &= 4U'_{12} + 12 \sum_{q}U'_2|_{Q_u\to Q_q},\  \nonumber \\
U'_{17} &\equiv \sum_q\hat{U}'_{17}= 2 \sum_{q}U_8|_{Q_u\to Q_q},\  
&U'_{18} &= U_8|_{Q_u\to Q_d},   \nonumber \\
U'_{19} &= \kappa Q_uQ_dU_8,\ 
&U'_{20} &= \sum_{q}\kappa Q_qQ_d\hat{U}'_{17},\  \nonumber \\
U'_{21} &= \kappa Q_d^2U'_{18} \, ,
\end{align}
with 
the functions $f_i(\s)$ given by
\begin{align}
f_1(\s)  =&\ \frac{\pi}{6}-\arctan(x) \, , \label{eq:f1}\\
f_2(\s)  =&\ 2 \, \text{Li}_2(-\s)-\ln^2(\s)+2 \, \ln(\s) \ln(1+\s)+\zeta_2 \, , \\
f_3(\s)  =&\ 4 \, \text{Li}_3(-\s)-2 \, \text{Li}_2(-\s) \ln(\s)-\frac{1}{6} \ln ^3(\s)+\zeta_2 \, \ln(\s)+3 \zeta_3 \, , \\
f_4(\s)=&\ 2 \text{Li}_2(1-\s)+\log ^2(\s)\, , \\
f_5(\s)=&\ -2 \text{Li}_3(\s)+\text{Li}_2(\s) \log (\s)+\zeta (2) \log (\s)+\frac{\log ^3(\s)}{6}+2 \zeta (3)\, , \\
f_6(\s)=&\ i\left[\text{Li}_2\left(\frac{2 (\s-1)}{\s-i \s/x-2}\right) - c.c. \right]\, ,\\
f_7(\s)=&\ 2 i {f_1}(\s) \left[\text{Li}_2\left(\frac{ 1 -i x   }{1+i x}\right)-c.c.\right] 
-\left[\text{Li}_3\left(\frac{ 1-i x }{1+i x}\right)+c.c.\right]+\frac{2 \zeta (3)}{3}\, ,\\
f_8(\s) =&\ 12 \text{Li}_2\left(-\sqrt{\s}\right)+\pi ^2, \\
f_{9}(\s)=&\ \pi ^2 + 4 \text{Li}_2\left(\sqrt{\s}-1\right)-8 \text{Li}_2\left(\sqrt{\s}\right) +2 \text{Li}_2(\s) \nonumber \\
&-4 \log\left(2-\sqrt{\s}\right) \log \left(\sqrt{\s}+1\right)-2 \log (1-\s) \log (\s), \\
f_{10}(\s)=& \log\left(1-\sqrt{\s}\right) \left(\pi -2 \arctan(x)\right)+4 \log \left(\sqrt{\s}+1\right) \arctan(x) \nonumber \\
&+i \left[ \text{Li}_2\left(\frac{2 x}{x-i}\right) + \text{Li}_2\left(\frac{2 x}{\sqrt{\s}(x+i)}\right) + \text{Li}_2\left(\frac{\s (x+i)}{2 (\s-1) x}\right)- c.c.\right] \nonumber \\
&+i \left[\text{Li}_2\left(\frac{2\left(\sqrt{\s} -1\right) x}{\sqrt{\s} (x+i)}\right) + \text{Li}_2\left(\frac{2 \left(\sqrt{\s}-1\right)x}{\s (x+i)-2 x}\right)- c.c. \right] .
\end{align}
We have defined $x\equiv \sqrt{\frac{\s}{4-\s}}$ to shorten the expressions. We emphasize that all the $f_i(\s)$ functions 
and also all the $U_i^{(')}(\s)$ functions are manifestly real.

\end{appendix}


\end{document}